\documentclass[prd,twocolumn,showpacs,preprintnumbers,floatfix,amsmath,amssymb,amsfonts]{revtex4}

\usepackage{graphicx}
\usepackage{amsmath}
\usepackage{dcolumn}
\usepackage{epsfig}
\usepackage{psfrag}
\usepackage{subfigure}
\usepackage{hyperref}

\RequirePackage{xspace}

\def\TeV{\ifmmode {\mathrm{\ Te\kern -0.1em V}}\else
	                   \textrm{Te\kern -0.1em V}\fi}%
\def\GeV{\ifmmode {\mathrm{\ Ge\kern -0.1em V}}\else
	                   \textrm{Ge\kern -0.1em V}\fi}%
\def\MeV{\ifmmode {\mathrm{\ Me\kern -0.1em V}}\else
	                   \textrm{Me\kern -0.1em V}\fi}%
\def\keV{\ifmmode {\mathrm{\ ke\kern -0.1em V}}\else
	                   \textrm{ke\kern -0.1em V}\fi}%
\def\eV{\ifmmode  {\mathrm{\ e\kern -0.1em V}}\else
	                   \textrm{e\kern -0.1em V}\fi}%
\let\tev=\TeV
\let\gev=\GeV

\def\TeVc{\ifmmode {\mathrm{\ Te\kern -0.1em V}/c}\else
	                   {\textrm{Te\kern -0.1em V}/$c$}\fi}%
\def\GeVc{\ifmmode {\mathrm{\ Ge\kern -0.1em V}/c}\else
	                   {\textrm{Ge\kern -0.1em V}/$c$}\fi}%
\def\MeVc{\ifmmode {\mathrm{\ Me\kern -0.1em V}/c}\else
	                   {\textrm{Me\kern -0.1em V}/$c$}\fi}%
\def\keVc{\ifmmode {\mathrm{\ ke\kern -0.1em V}/c}\else
	                   {\textrm{ke\kern -0.1em V}/$c$}\fi}%
\def\eVc{\ifmmode  {\mathrm{\ e\kern -0.1em V}/c}\else
	                   {\textrm{e\kern -0.1em V}/$c$}\fi}%
	
\def\cm{\ifmmode  {\mathrm{\ cm}}\else	                   
\textrm{~cm}\fi}%

\def\mm{\ifmmode  {\mathrm{\ mm}}\else	                   
\textrm{~mm}\fi}%
	
\def\ifb{\mbox{fb$^{-1}$}}%  Inverse femtobarns.
\usepackage{relsize}
\def\babar{\mbox{\slshape B\kern-0.1em{\smaller A}\kern-0.1em
    B\kern-0.1em{\smaller A\kern-0.2em R}}}

\begin{document}
%\linenumbers

\title{Identification of a bottom quark-antiquark pair in a single jet \\with high transverse momentum and its application}

\author{Chunhui Chen} 
\affiliation{Department of Physics and Astronomy, Iowa State University, Ames, Iowa 50011, USA}

\begin{abstract}
In this paper we introduce a new approach to identify  a bottom quark-antiquark pair  inside a single jet 
with high transverse momentum by using the jet substructure in the center-of-mass frame of the
jet. We demonstrate that the method can be used to discriminate the boosted heavy particles 
decaying to a $b\bar{b}$ final state from QCD jets.  Applications to searches for 
the standard model Higgs boson ($H$) decaying to $b\bar{b}$ when produced in association
with a weak vector boson are also discussed.
\end{abstract}

\maketitle

\section{Introduction}
\label{sec:intro}
The existence of a Higgs boson-like particle with a mass of around 125\,\gev\  
has been firmly established by both the ATLAS and CMS experiments~\cite{Aad:2012tfa,Chatrchyan:2012xdj} in its bosonic 
decay modes ($H\to\gamma\gamma$, $H\to ZZ$, and $H\to WW$). However, its decay to  a 
bottom quark-antiquark pair ($b\bar{b}$) final state has not been observed yet. 
Finding such a decay signal at the LHC  is challenging because 
of the large amount background from the production of multijets containing $b$ quarks, despite that
the $H\to b\bar{b}$ decay mode is predicted in the standard model (SM) to have a branch fraction of $58\,\%$
for $m_H=125\,\GeV$. As a result, 
such searches  have been mostly performed  in the $pp\to VH$ production mode, where 
$V$ is either a $W$ or a $Z$ boson that decays leptonically, and $H\to b\bar{b}$. So far, no evidence 
of such a decay has yet been seen 
by either the ATLAS or CMS experiment~\cite{Chatrchyan:2013zna,Aad:2014xzb}.

It has been shown that the search sensitivity of $H\to b\bar{b}$ in the $VH$ production mode can be
significantly improved by reconstructing the hadronically decaying Higgs boson with large transverse momentum 
in a single jet~\cite{Butterworth:2008iy},  especially together with the implementation of
the jet substructure techniques~\cite{Soper:2010xk,Kim:2010uj,Butterworth:2015bya}. Such an approach requires the identification of both $b$ quarks
decaying from the Higgs boson in a single jet, hereafter referred to as a Higgs jet ($H$ jet). While the identification of an isolated jet stemming from the hadronization 
of a single $b$ quark ($b$-tagging) has been widely used in many experimental measurements,  its application to the 
Higgs jets is not trivial since  a Higgs jet has two $b$ quarks inside~\cite{ATLAS-PUB-2014-013,ATLAS-PUB-2015-035}. In this  paper, 
we extend the studies presented in Refs.~\cite{Chen:2011ah,Chen:2013ola,Chen:2013uja} to explore the identification of
the $b\bar{b}$ pair inside a $H$ jet (double $b$-tagging) in the center-of-mass frame of the jet. 
We demonstrate that the method can greatly reduce the QCD jet background while maintaining a high
identification efficiency of the boosted Higgs boson even in an environment with very large numbers
of multiple interactions per event (pileup), where the QCD jets are defined as those jets initiated by a nontop quark or gluon.

We organize this paper as follows: In Sec.~\ref{sec:sample}, we describe the event sample we used  in the study.
Section~\ref{sec:jet_sub} discusses the method to identify  a bottom quark-antiquark pair  in a single jet in the jet center-of-mass frame and its performance.
Applications of our method to the searches of $H\to b\bar{b}$ in the $VH$ production mode
are discussed in Sec.~\ref{sec:app}. We conclude in Sec.~\ref{sec:conclusion}.

\section{Event Sample}
\label{sec:sample}
We use the boosted $H$ jets, from the SM process of  $WH$  production,  as an benchmark to 
illustrate our proposed double $b$-tagging method. For simplicity, we only consider the background from 
the SM $W$+jets production to study the background rejection performance of the QCD jets
as it is the largest background in searches for $H\to b\bar{b}$ in the $WH$ production mode.
However, our method is generic and is applicable to any boosted heavy particles decaying to a $b\bar{b}$ final state.
In addition, we also generate events to simulate the SM processes of  
$ZH$, $WZ$, $WW$, $ZZ$, $Z$+jets and top quark production.

All the events used in this analysis are produced using the P{\footnotesize ythia} 8.186 Monte Carlo (MC) event generator~\cite{Sjostrand:2006za,Sjostrand:2007gs}
for the $pp$ collision at $14\,\rm TeV$ center-of-mass energy.
To simulate the finite resolution of the
Calorimeter detector at the LHC experiments, we divide the $(\eta, \phi)$ plane into $0.1\times 0.1$ cells. 
The energies of particles entering each cell in each event, except for the neutrinos, are summed over and
replaced with a massless pseudoparticle of the same energy, also referred as an energy cluster,  pointing to the center of the  cell. 
These pseudoparticles are fed into the 
F{\footnotesize astJet} 3.0.1~\cite{Cacciari:2005hq} package for  jet reconstruction.
The jets are reconstructed with the anti-$k_T$ algorithm~\cite{Cacciari:2008gp}  
with a distance parameter of $\Delta R=0.8$. The  anti-$k_T$ jet algorithm is the default one used at the ATLAS and CMS experiments. 
As for the charged tracks, their momentum and vertex positions are smeared according to the expected resolutions of the ATLAS detector~\cite{Aad:2008zzm}.
To evaluate the performance of the double $b$-tagging with the currently expected experimental conditions at
the LHC, we generate the MC events with different average numbers of multiple interactions per event, where
the beam spot is assumed to follow a Gaussian distribution with a width of $0.015\,\mm$ in the transverse beam direction, and 
$45\,\mm$ in the longitudinal beam direction.
We then perform our studies for each scenario and compare their performances. 

\section{Double $b$-tagging and Jet Substructure in the rest frame}
\label{sec:jet_sub}
\subsection{Event selection}
In this section we describe the method to identify  a bottom quark-antiquark pair in a single jet 
using the substructure in the center-of-mass frame of
the jet in order to distinguish a boosted hadronically decaying Higgs boson from QCD jets. 

The study is done using the MC simulated events of the $WH$ and $W$+jets productions, where 
the $W$ boson decays leptonically ($W\to e\nu, \mu\nu$). We select events with one isolated lepton (electron or muon)
with $p_{\rm T}>20\,\gev$ and $|\eta|<2.4$, where $p_{\rm T}$ and $\eta$ are the transverse momentum and psudorapidy
of the lepton.  For the jet reconstruction, studies show that its energy and invariant mass ($m_{\rm jet}$) can significantly 
shift to higher values due to the presence of additional energy depositions from underlying events and pileups. 
We employ a jet area correction technique~\cite{Cacciari:2007fd}
to take into account the effects  on the event-by-event basis.
For each event, a distribution of transverse energy densities is calculated for all jets with $|\eta|<2.1$, and its median is taken as
an estimate of the energy density of  the pileup and underlying events. 
We subsequently correct each jet by subtracting the product of the transverse energy density and the jet area, which is
determined with the ``active" area calculation technique~\cite{Cacciari:2007fd}. This method results in a modified jet four-momentum
that is used throughout the paper unless explicitly stated otherwise.
The jets with  $p_{\rm T}\ge\,300\,\gev$, $|\eta|\le1.7$, and $40\,\gev\le m_{\rm jet}\le 240\,\gev$
in an event are selected as the $H$ jet candidates for further analysis.

For $b$-tagging, only charged tracks with $p_{\rm T} >1\,\gev$ and $|\eta|<2.5$ are considered. They
are also required to satisfy the criteria that  $|d_0|<1\,\mm$ and $|z_0-z_{\rm pv}|\sin\theta<1.5\,\mm$,
where $d_0$ and $z_0$ are the transverse and longitudinal impact parameters of the charged track, 
$z_{\rm pv}$ is the longitudinal position of the primary vertex, and $\theta$ is the polar angle of the
charged track. A  charged track is considered to be associated with a jet if the distance parameter of 
$\Delta R=\sqrt{\Delta\eta^2+\Delta\phi^2}$ is less than $0.8$, where $\Delta\eta$ and $\Delta\phi$ are 
defined as the differences in psudorapidity and the azimuthal angle between the charged track and the jet,
respectively.

\subsection{Center-of-mass  frame of  a jet}
\begin{figure}[!htb]
\begin{center}
\includegraphics[width=0.45\textwidth]{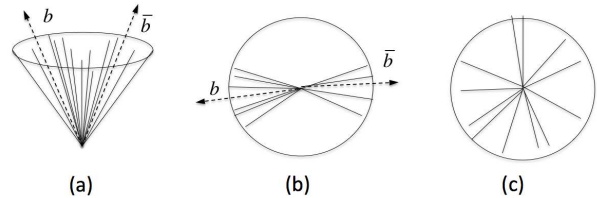}
\caption{Illustration of the constituent particle distribution of a jet. (a) $H$ jet in the lab frame. (b) $H$ jet in the jet rest  frame. 
 (c) QCD jet in its rest frame. }
\label{fig:Bst_scheme}
\end{center}
\end{figure}

We define the center-of-mass frame (rest frame) of a jet as the frame where the four-momentum of the
jet is equal to $p^{\rm rest}_{\mu}\equiv (m_{\rm jet}, 0, 0, 0)$. A jet consists of its constituent particles.
The distribution of the constituent particles of a boosted  Higgs jet in its center-of-mass frame
looks like a back-to-back dijet event with one $b$ quark in each of the subjets. On the other hand, a QCD jet acquires its mass through
gluon radiation and it is not a closed system. The constituent particle distribution of a QCD jet in the rest frame  is more likely to be random, as illustrated in Fig.~\ref{fig:Bst_scheme}. 

\subsection{Reclustering}
We recluster the energy clusters of a jet to reconstruct subjets in the jet rest frame using a modified 
$e^+e^-$ Cambridge jet reconstruction algorithm~\cite{Dokshitzer:1997in} in 
the F{\footnotesize astJet} 3.0.1~\cite{Cacciari:2005hq} package 
by replacing the distance parameter with a new choice of the distance parameter, $\Omega$,
where $\Omega$ is defined as the angle between two pseudoparticles in the jet rest frame.
The algorithm performs a sequential recombination of the pair of psedoparticles that is closest in angle $\Omega$,
except for $\Omega > 0.8$. The reconstructed subjets  are
required to have energy $E_{\rm subjet}>10\,\gev$ in the $H$ jet rest frame. 
We then boost all the tracks associated with the $H$ jet candidate back to the center-of-mass frame of the jet.  A charged track is considered 
to be associated with a subjet only if their angular separation is less than 0.8 in the jet rest frame. By doing so, we separate the charged tracks 
that originate from different partons of the Higgs boson decay  and reject many charged tracks from underlying events and pileup. 
This allows a straightforward identification of the $b$ quarks inside the $H$ jets by applying the existing $b$-tagging algorithms
on the charged tracks associated with each subjet. In our analysis, we only retain the jets if their two subjets with the highest energy  (leading subjet)
have at least one associated charged track. Those two subjets are considered as the $b$ and $\bar{b}$ subjet candidates of the
$H$ jet.

\subsection{Double $b$-tagging}
\label{sec:doube-btagging}
\begin{figure*}[!htb]
\begin{center}
\includegraphics[width=0.23\textwidth]{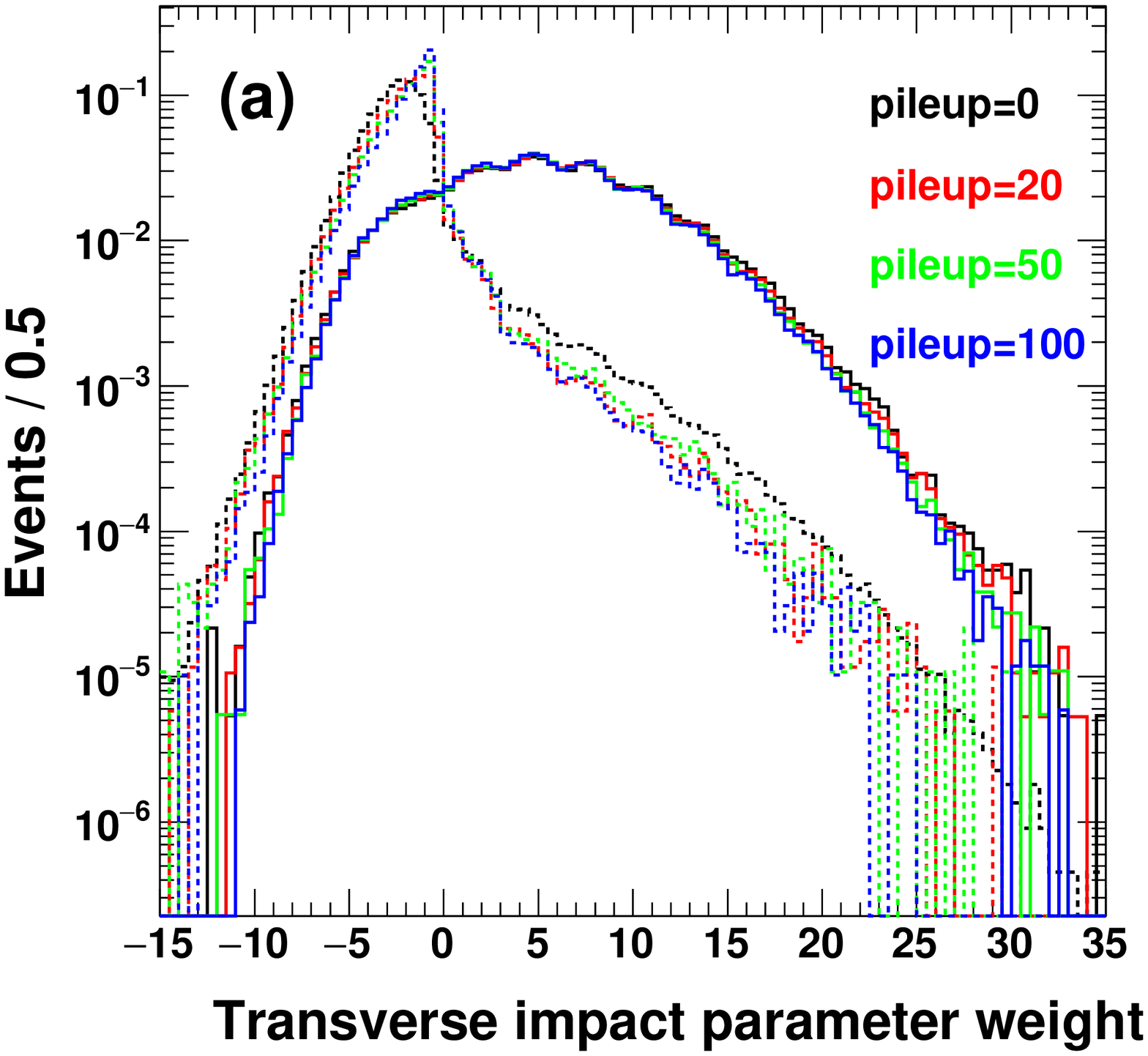}
\includegraphics[width=0.23\textwidth]{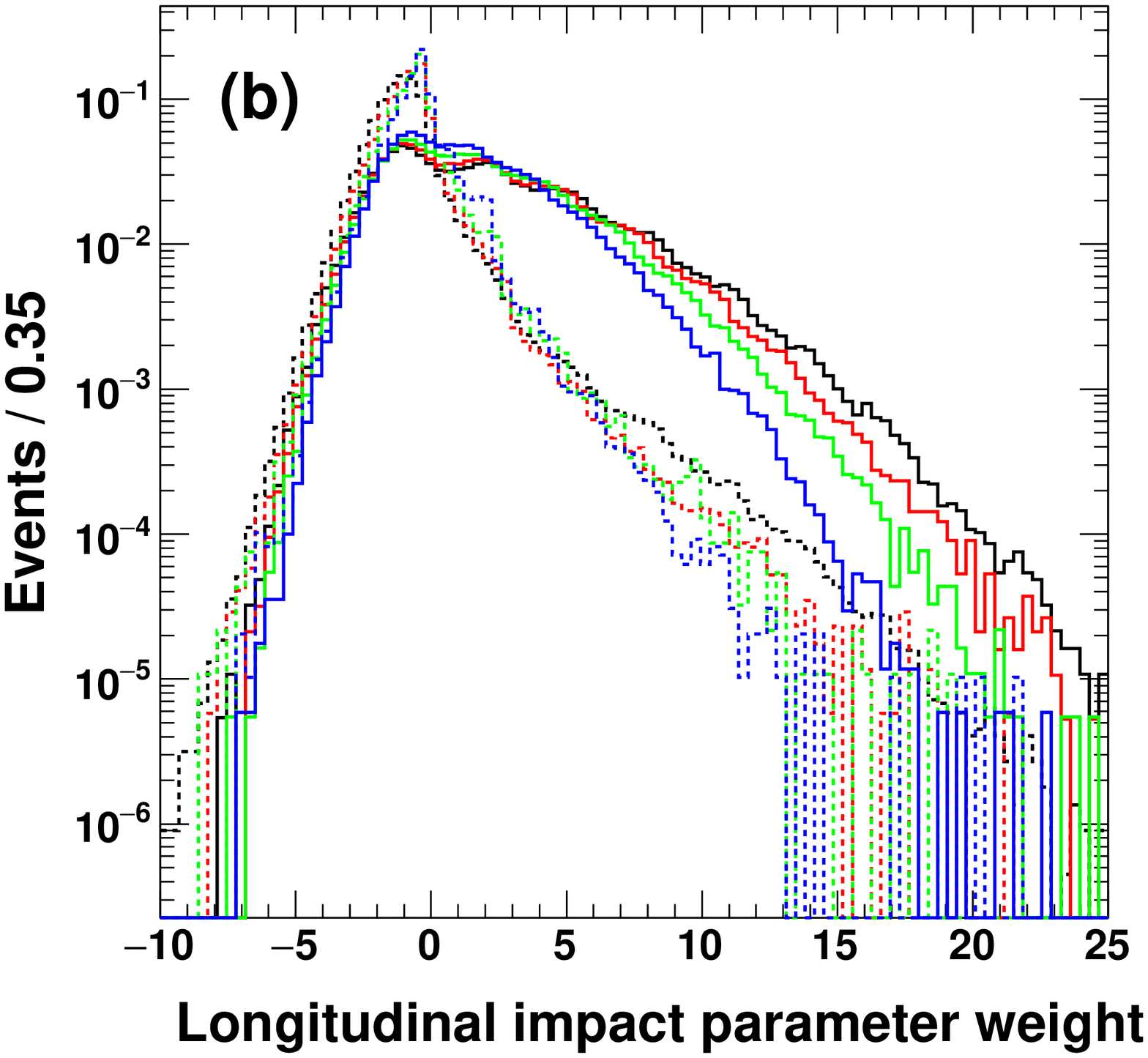}
\includegraphics[width=0.23\textwidth]{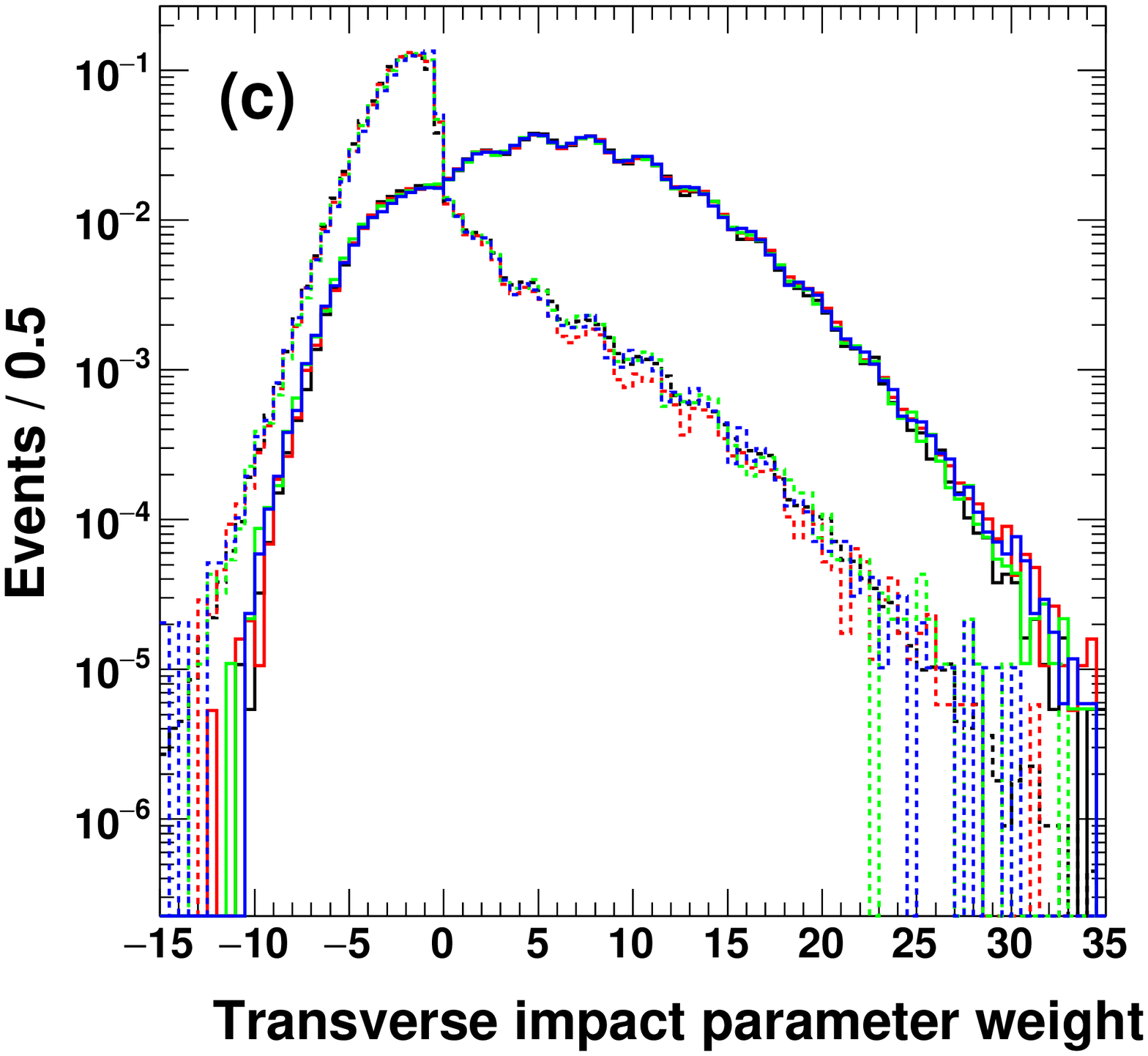}
\includegraphics[width=0.23\textwidth]{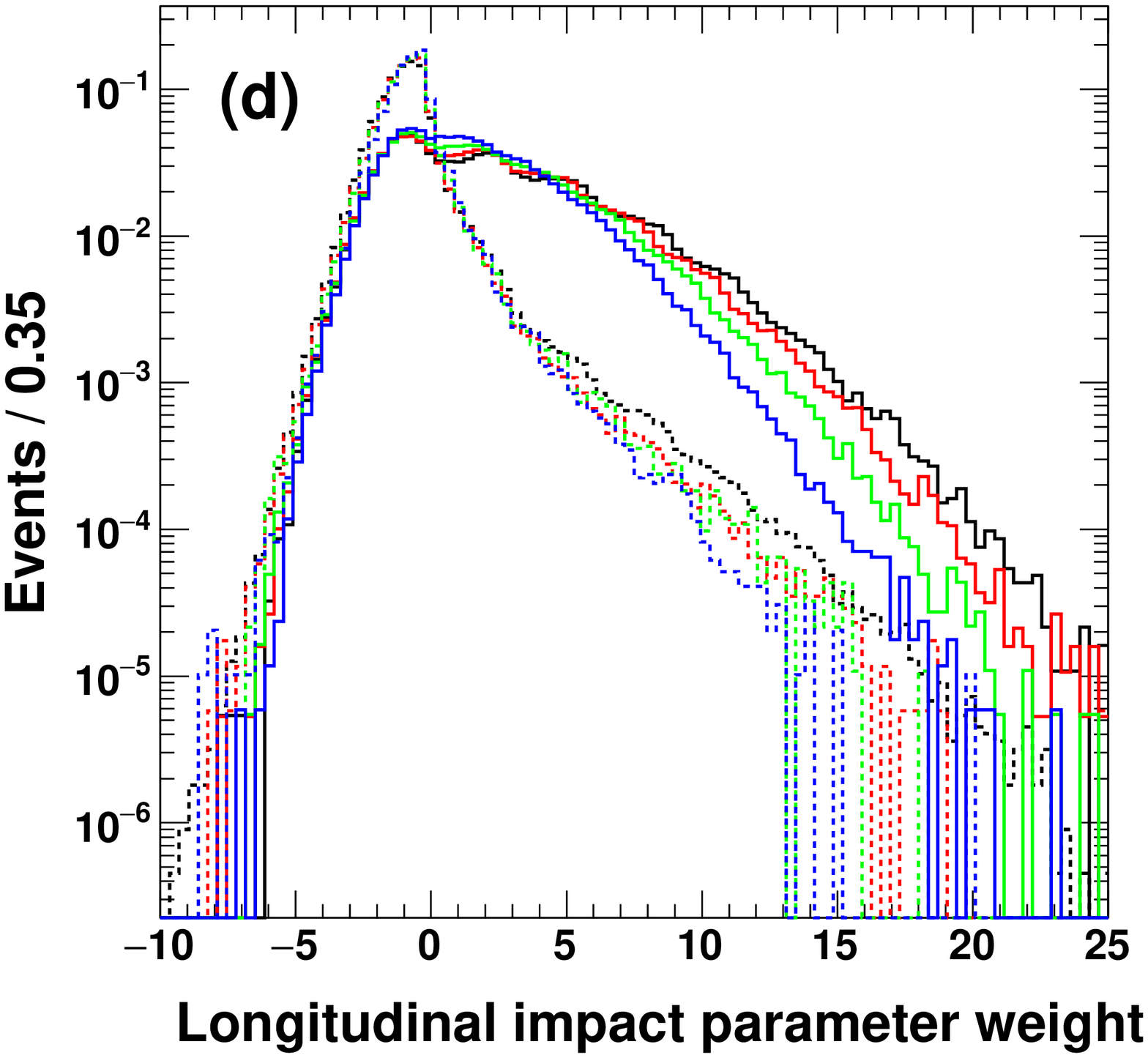}
\caption{The subjet weight distributions of the transverse and longitudinal impact parameter weight
under different pileup conditions. In (a) and (b), the solid (dashed) lines represent the distributions of the charged  
tracks associated with the subjets that have the highest energy in the $H$ (QCD) jet rest frame.
In (c) and (d), the solid (dashed) lines represent the distributions of the charged  
tracks associated with the subjets that have the second-highest energy in the $H$ (QCD) jet rest frame.
All the distributions are normalized to unity.}
\label{fig:IPweight}
\vspace{0.5cm}
\includegraphics[width=0.23\textwidth]{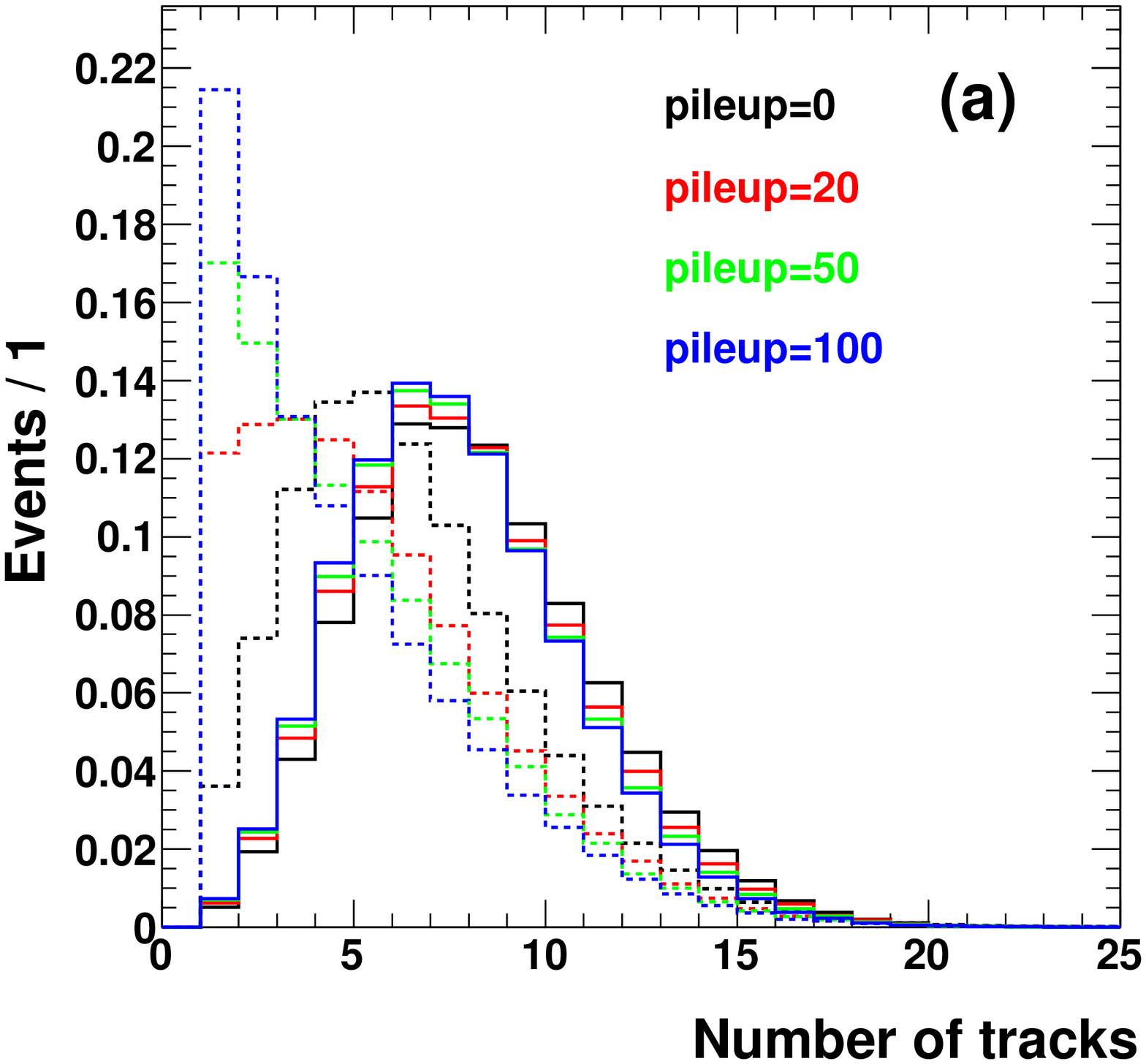}
\includegraphics[width=0.23\textwidth]{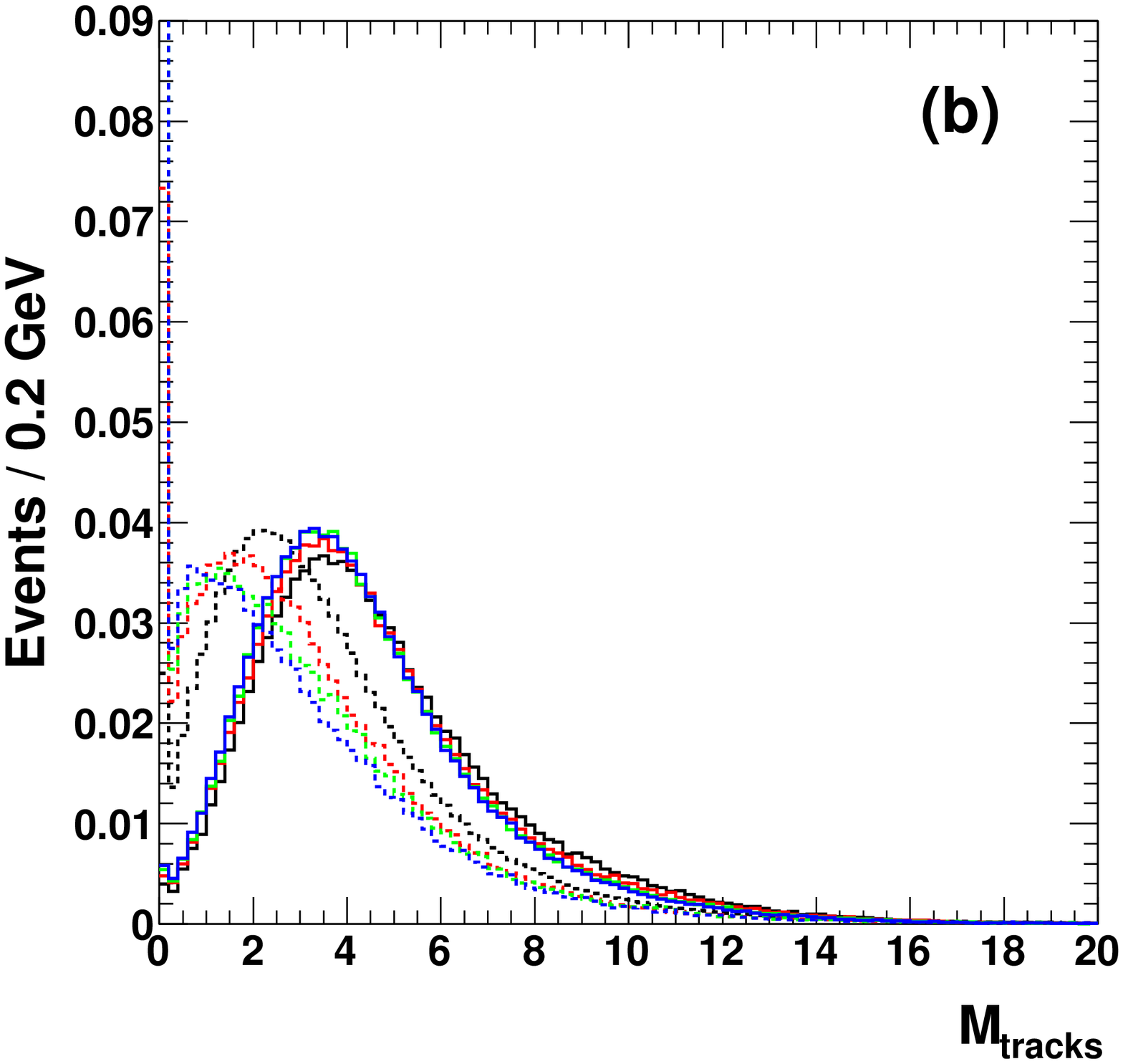}
\includegraphics[width=0.23\textwidth]{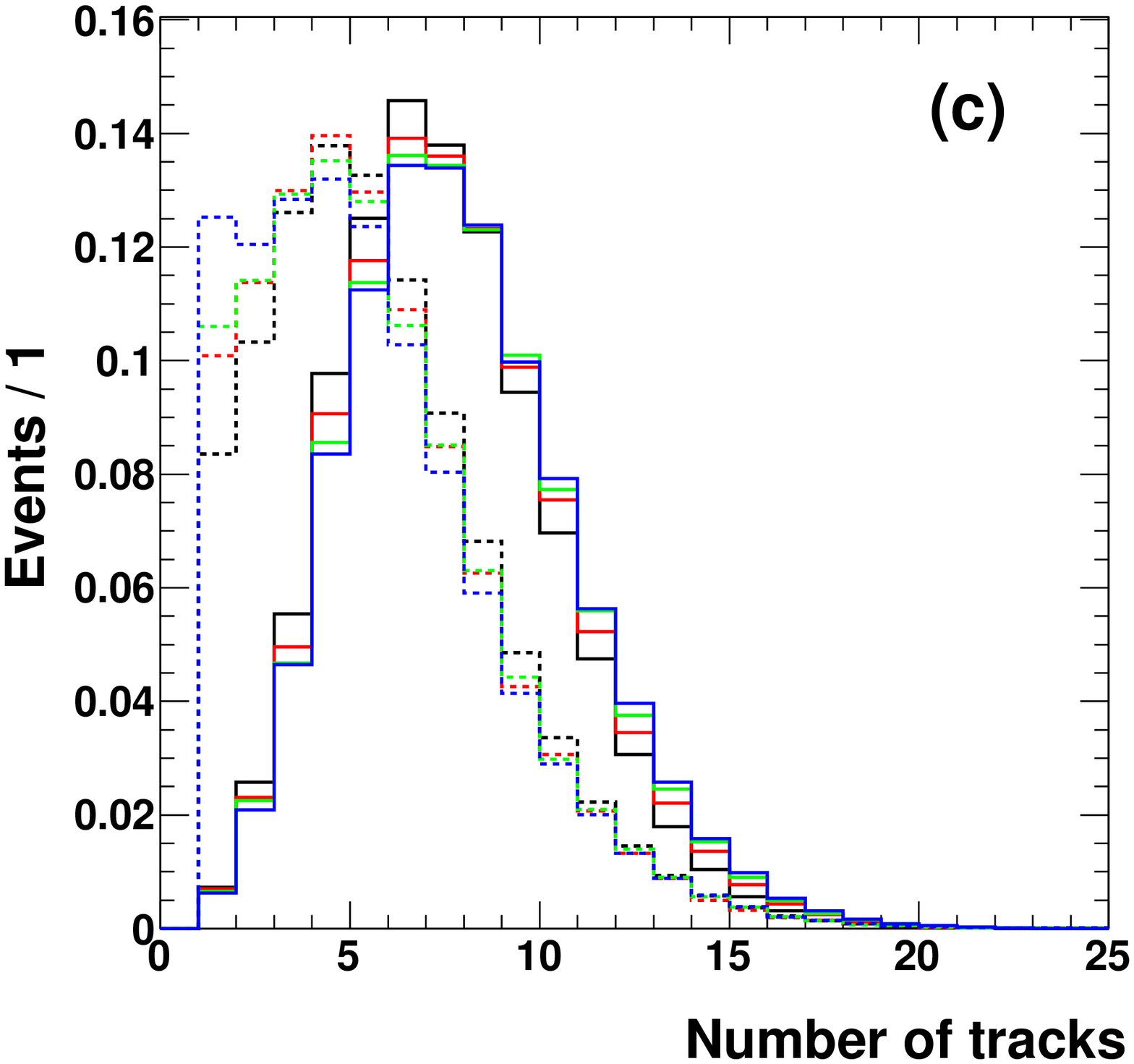}
\includegraphics[width=0.23\textwidth]{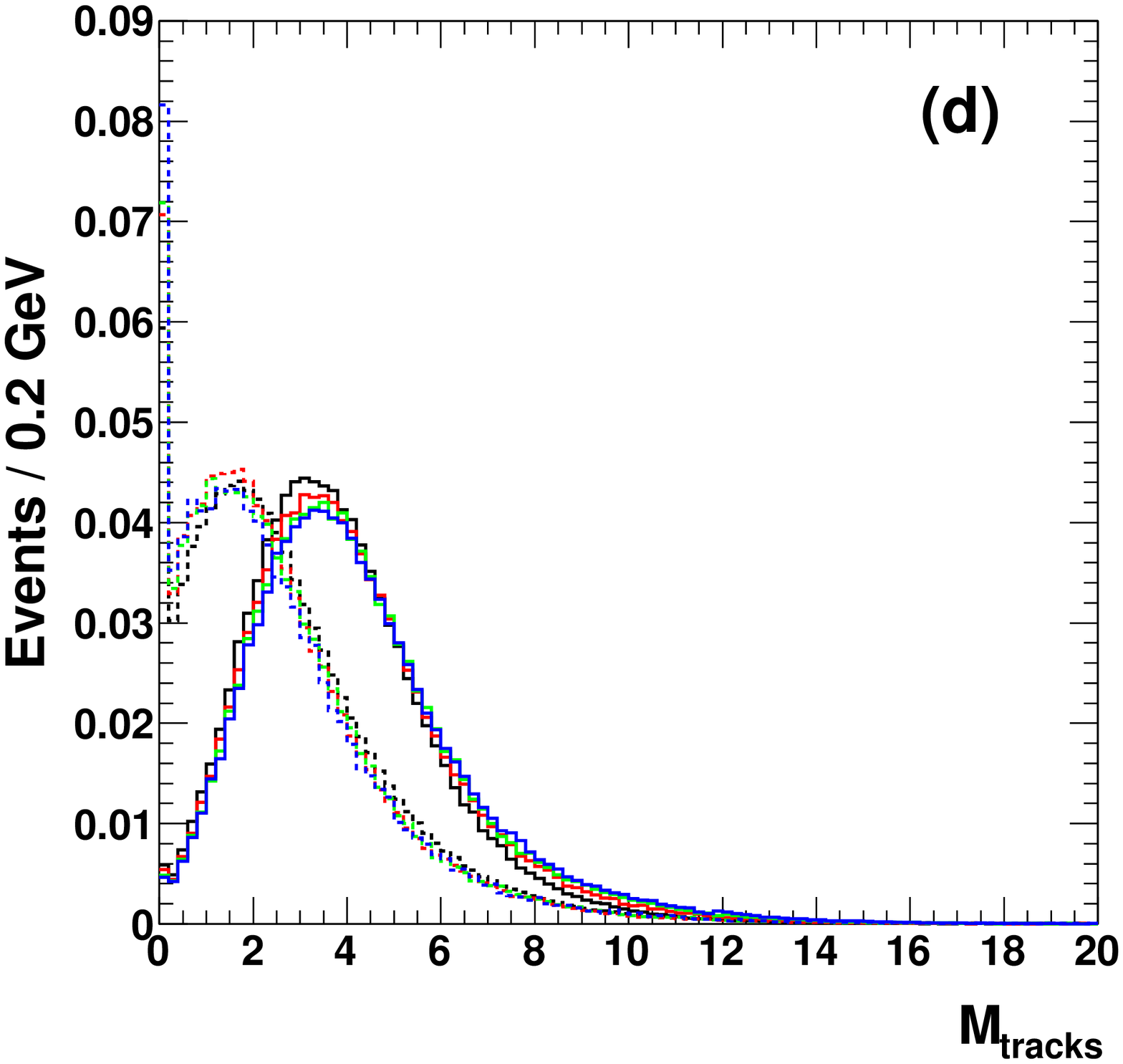}
\caption{The distribution of the number and invariant mass of the charged tracks that are associated
with the subjets in the rest frame of the $H$ (solid line)  and QCD (dashed line) jets under different pileup conditions.
In (a) and (b), the distributions are from the subjets with the highest energy in the jet rest frame.
In (c) and (d), the distribution are from the subjets with the second highest energy in the jet rest frame.
All the distributions are normalized to unity.}
\label{fig:IPtrk}
\end{center}
\end{figure*}

In this paper, we illustrate the double $b$-tagging in the jet rest frame with a tagging algorithm based on the 
charged track impact parameters since this algorithm is widely used in many experiments and is easy to implement.
It is also among the official $b$-tagging methods used by the ATLAS experiment~\cite{Aad:2009wy}. 
The impact parameters of tracks are computed with respect to the primary vertex in the lab frame. They typically have significant nonzero values for the charged tracks
from the $b$-hadron decays because of its long lifetime. The impact parameter is signed to further discriminate the
tracks from $b$-hadron decay from tracks originating from the primary vertex based on the fact that the decay position 
of the $b$ hadron lies along its flight path. The sign of transverse impact parameter $d_0$ is determined using the
subjet momentum $\vec{p}_{\rm subjet}$, the track momentum $\vec{p}_{\rm trk}$ at the point of the closest approach $\vec{x}_{\rm trk}$~\cite{Aad:2009wy}
to the primary vertex position $\vec{x}_{\rm pv}$:
\begin{equation}
{\rm sign}(d_0)=(\vec{p}_{\rm subjet}\times \vec{p}_{\rm trk})\cdot (\vec{p}_{\rm trk}\times (\vec{x}_{\rm pv}-\vec{x}_{\rm trk})).
\end{equation}
The sign of longitudinal impact parameter $z_0$ is measured by the sign of $(\eta_{\rm subjet}-\eta_{\rm trk})\times z_{0,{\rm trk}}$,
where $\eta_{\rm subjet}$ is the pseudorapidity of the subjet, and $\eta_{\rm trk}$ and $z_{0,\rm trk}$ are the
pseudorapidity and longitudinal impact parameters of the track at the position $\vec{x}_{\rm trk}$, respectively.
All the quantities in the computation of the signed impact parameters are the ones defined in the lab frame.

We form a likelihood of the charged tracks associated with a subjet. The measured impact parameter significance $S_i$
of the $i$th track in a subjet is compared to the predefined functions for both $b$ subjet and non-$b$ subjet hypothesis, $b(S_i)$
and $u(S_i)$, where $b(S)$ and $u(S)$ are the smoothed and normalized distributions of the charged tracks
that are associated with the $b$ subjets in the signal $H$ jets and the subjets in the QCD jets, respectively.
The ratio of the probabilities $b(S_i)/u(S_i)$ defines a weight $W_i$. A subjet weight $W_{\rm subjet}$ is then calculated 
as the sum of the $W_i$ from all the charged tracks associated with the subjet. In case there are no charged tracks associated with a subjet,
its subjet weight is assigned to be zero.  
The distributions of the subjet weights of the $H$ and QCD jets are shown in Fig~\ref{fig:IPweight}. It shows a clear separation between the signal and background distributions.

To  further help identify subjets that are originated from a $b$ quark, we explore two
additional properties of the subjet: the number and the invariant mass of 
the charged tracks that are associated with the subjet. Their distributions of the $H$ and QCD jets
are shown in Fig.~\ref{fig:IPtrk}. 
\begin{figure*}[!htb]
\begin{center}
\includegraphics[width=0.46\textwidth]{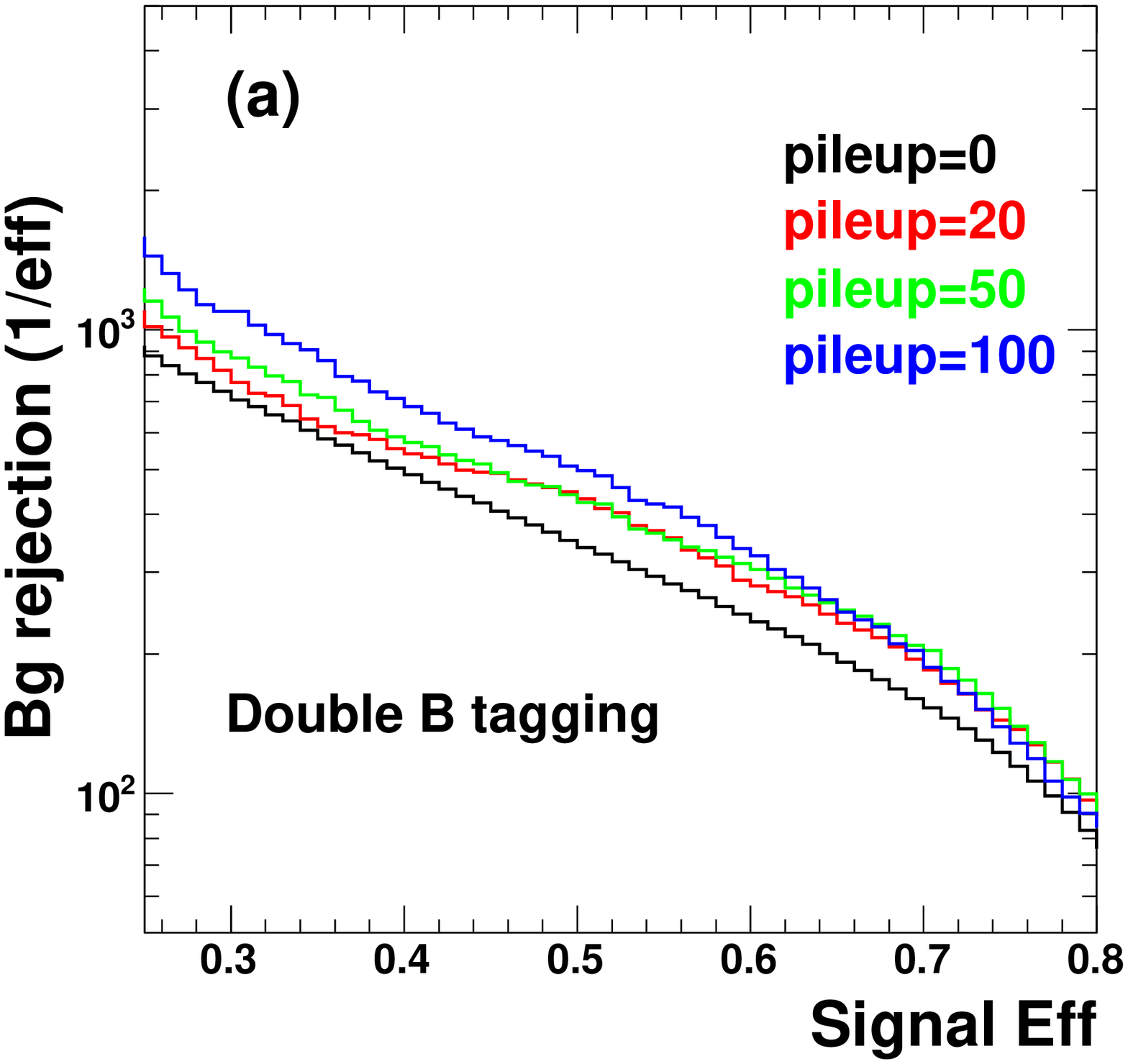}
\includegraphics[width=0.46\textwidth]{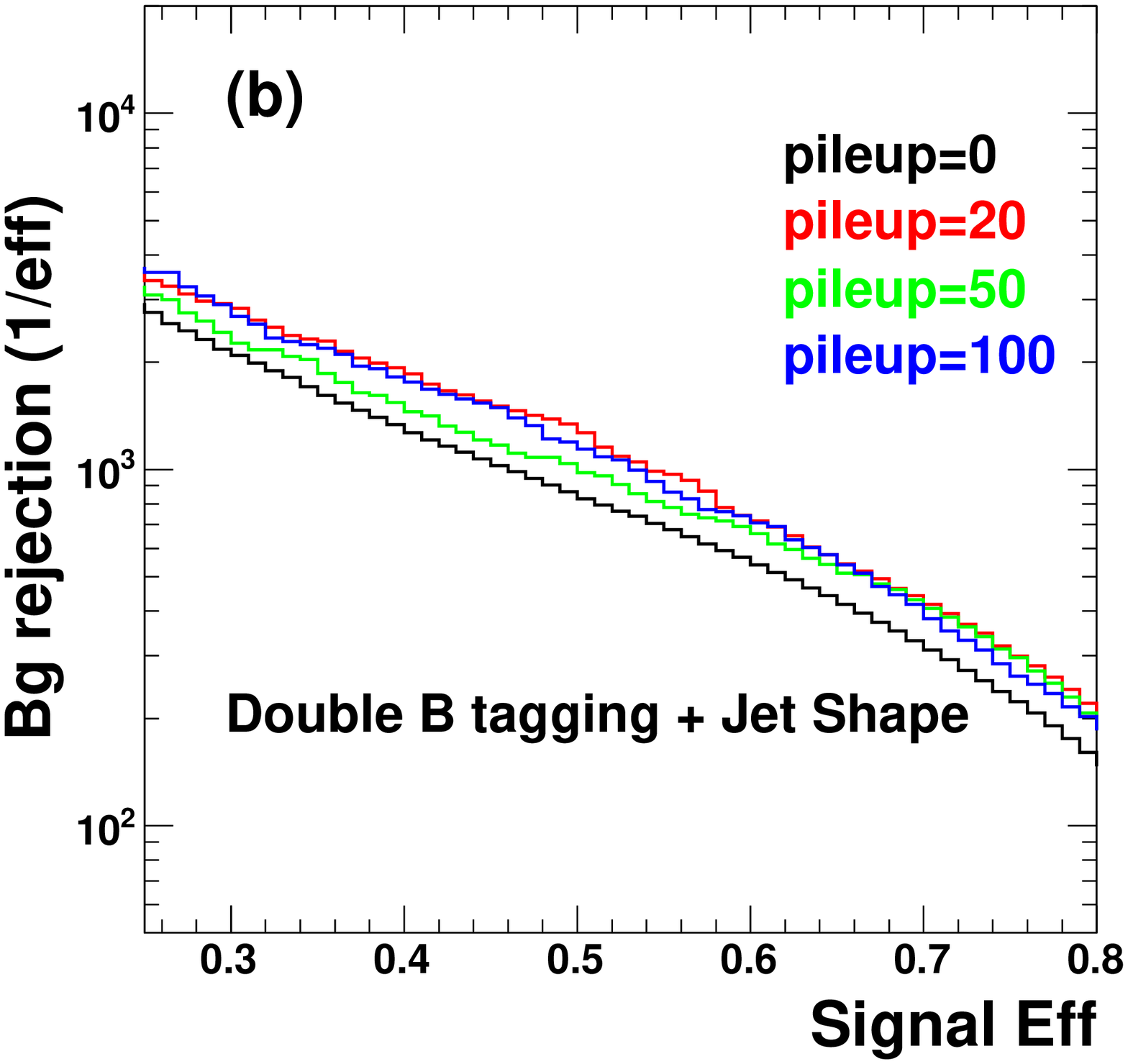}
\caption{The background rejection of QCD jets vs. the signal efficiency of $H$ jets in different
pileup conditions based on the double $b$-tagging without (a) and with (b) using the jet substructure
information in its rest frame.}
\label{fig:SigEff_BgRej_btag}
\end{center}
\end{figure*}

The final double $b$-tagging variable is constructed using a boosted decision tree (BDT) algorithm with the subjet weights of the first two
leading subjets in the jet rest frame, the numbers and invariant masses of the charged tracks associated with the first two leading subjets. 
The signal efficiency of $H$ jets by identifying bottom quark and antiquark inside vs the background rejection of QCD jets for the BDT 
variable is shown in Fig~\ref{fig:SigEff_BgRej_btag}. Note that the performance of the double $b$-tagging is slightly better with higher pileup
at certain signal efficiencies. This is an effect that is caused by the selection of the jets used in the evaluation of the double 
$b$-tagging performance. In our analysis, we only use jets that have $p_{\rm T}>300\,\gev$, $40\le m_{\rm jet}\le250\,\gev$, and at least two subjets with 
$E_{\rm subjet}>10\,\gev$ and nonzero charged tracks in its rest frame. As a result, when the pileup increases, some QCD jets that otherwise 
would not pass the jet selection criteria can be selected. We repeat the studies of double $b$-tagging for jets selected with a higher $p_{\rm T}$ 
requirement (up to 1\,\tev) and observe no degrading of the performance. For a given signal efficiency, the background rejection is actually slightly higher
for jets with higher transverse momentum. This is primarily due to the fact that the displaced decaying vertices of $b$ hadrons in higher $p_{\rm T}$ jets are
further away from the beam spot as they have larger Lorentz boosts.

\subsection{Jet substructure}
\label{sec:tag_shape}
The jet substructure information can be used to improve the identification of the boosted $H\to b\bar{b}$ in addition of the
double $b$-tagging.  Here we demonstrate it by combining the double $b$-tagging with the
jet substructure variables defined in the jet rest frame, introduced in Ref.~\cite{Chen:2011ah}. They are
the thrust, thrust-minor, sphericity, aplanarity, and Fox-Wolfram Moments $R_2$. Those variables are designed to identify
a boosted two-body decay heavy particle of which the final decaying products are reconstructed in a single jet.
They have been successfully implemented by the ATLAS experiment to make the first observation of the boosted hadronically decaying vector boson 
reconstructed as a single jet from the SM $W/Z$+jets production~\cite{Aad:2014haa}. In addition, we introduce another variable $\cos\Theta$, where $\Theta$ is defined as the angle between the direction 
of the thrust axis of a jet in its rest frame~\cite{Chen:2011ah} and the jet momentum direction. 

We form a BDT variable using the jet substructure variables described above with the variable used in the double $b$-tagging 
in Sec.~\ref{sec:doube-btagging}.
Studies show that the jet substructure variables calculated based on the energy clusters have a great dependence on the
pileup condition because of the additional  energy depositions from the pileup and underlying events. 
To minimize the effect, the jet substructure variables in this paper are all computed using the charged tracks associated with the jet.
Their distributions for the $H$ jets and the QCD jets under different pileup conditions 
can be found in Appendix~\ref{appx:1}.  As shown in Fig.~\ref{fig:SigEff_BgRej_btag}, the background rejection
achieved by combining the double $b$-tagging and the jet substructure variables in the jet rest frame is a factor of 2 to 3  better compared to the one that
only relies on the double $b$-tagging.

\section{Application}
\label{sec:app}
In this section, we study two examples of the application of the double $b$-tagging algorithm in
searches for $H\to b\bar{b}$ in the $VH$ production modes, where the $W/Z$ boson decays
leptonically. For simplicity, we only consider the
kinematic region of the $VH$ production where the Higgs boson has a relatively high transverse momentum 
so that its hadronically decaying products can be reconstructed in a single jet. In both examples, we assume that 
the average pileup at the LHC is 50.

\subsection{Search for $H\to b\bar{b}$ in the $WH$ production mode}
In this search channel, the  leptonically decaying $W$ boson is reconstructed by requiring exactly one isolated lepton with $p_{\rm T}>20\gev$, $|\eta|<2.5$
and more than $25\gev$ of missing transverse energy ($E^{\rm miss}_{\rm T}$) in an event. We then select jets with $p_{\rm T}>300\,\gev$, $|\eta|<1.7$ 
and $40\le\,m_{\rm jet}\le 240\,\gev$ in the event as the 
hadonically decaying Higgs boson candidates. The jet is reconstructed with the anti-$k_{\rm T}$ algorithm with a distance parameter
$\Delta R=0.8$. To reduce the large amount of background from the SM $W$+jets production, a selection on
the BDT variable based on the double $b$-tagging and the jet substructure information as described in Sec.~\ref{sec:tag_shape} is applied. We optimize the 
selection cut on the BDT variable by maximizing $S/\sqrt{B}$, where $S$ and $B$ are the numbers of the signal and background events within $20\,\gev$ of the 
Higgs boson mass. In addition, we reject an event if it has a $b$ jet that is not overlapping with the selected $H$ jet candidate.  
This selection significantly reduces the background from the SM $t\bar{t}$ production. The jet mass distribution of the $H$ jet candidates after
all the event selection is applied is shown in Fig.~\ref{fig:mWHbb}. The significance of $S/\sqrt{B}$ in the signal window is about $4$ assuming 
$400\,\ifb$ of LHC data at the $14\,\tev$ center-of-mass energy.

\begin{figure}[htb]
\begin{center}
\includegraphics[width=0.48\textwidth]{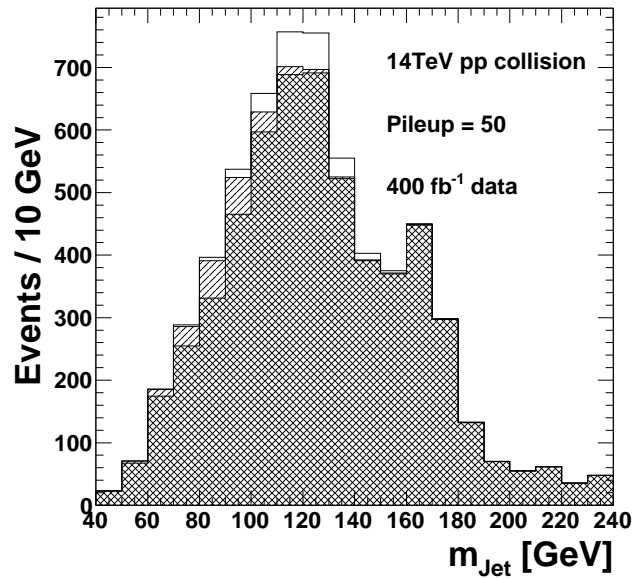}
\caption{Jet mass distribution of the selected $H$ jet candidates 
in the MC simulated event sample that is equivalent to $400\,\ifb$ of LHC data at the $14\,\tev$ center-of-mass 
energy after all the event selection criteria are applied. The open histogram represents the expected contribution of the $WH$ signal events.
The left hashed histogram represents the expected contribution of the $WZ$ production, where $Z\to b\bar{b}$ is
also reconstructed as a single jet.
The right hashed histogram shows the expected background that is dominated by the top productions ($>80\,\%$)
with also a significant contribution from the $W$+jets production.
The peaking structure around $160\,\gev$ is from the hadronically decaying top quark from the SM top production.}
\label{fig:mWHbb}
\end{center}
\end{figure}

\subsection{Search for $H\to b\bar{b}$ in the $ZH$ production mode}
\begin{figure}[!htb]
\begin{center}
\includegraphics[width=0.48\textwidth]{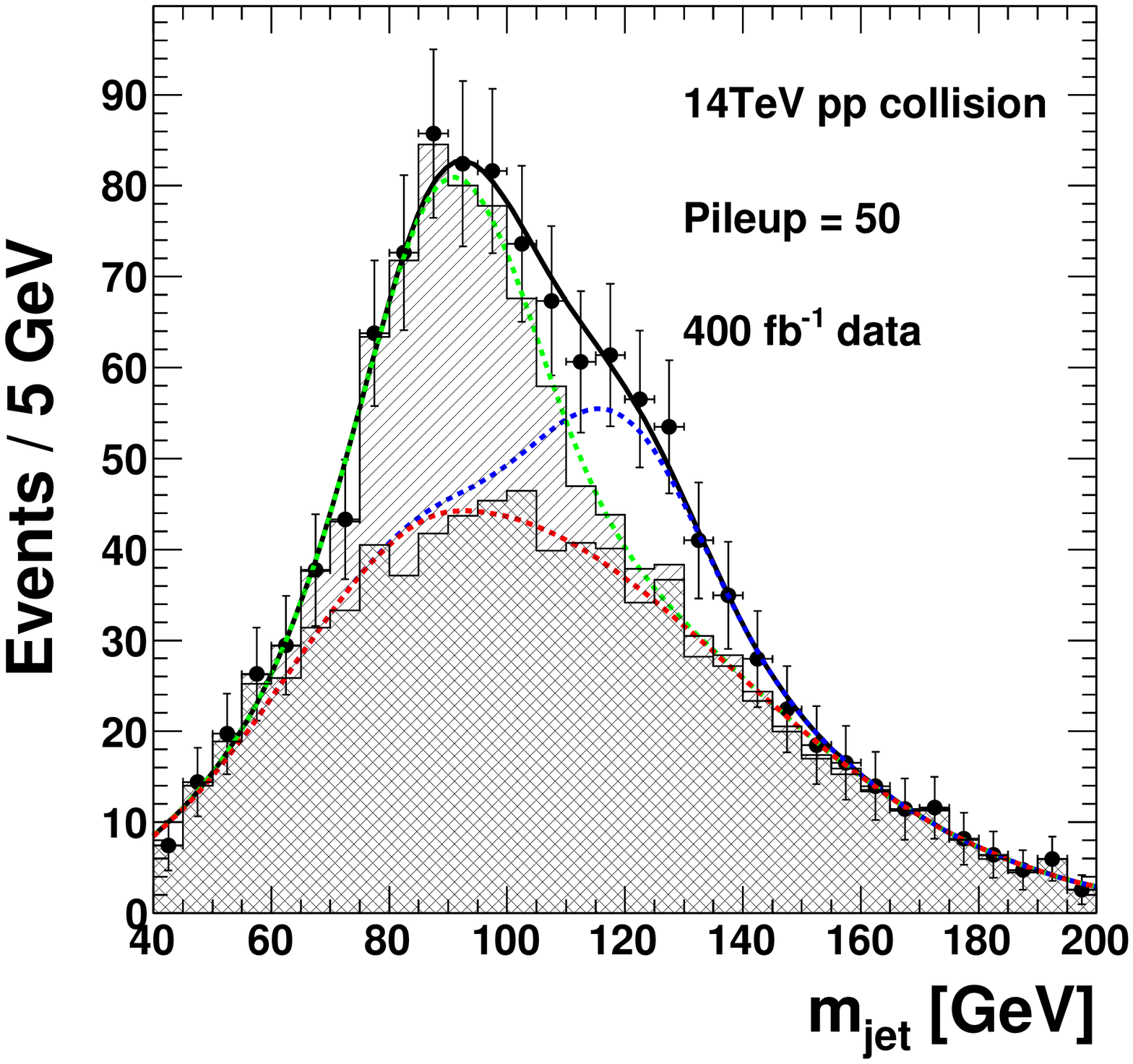}
\caption{Jet mass distribution of the selected $H$ jet candidates 
in the MC simulated event sample that is equivalent to $400\,\ifb$ of LHC data at the $14\,\tev$ center-of-mass 
energy after all the event selection criteria are applied. The open histogram represents the expected contribution of the $ZH$ signal events.
The left hashed histogram represents the expected contribution of the $ZZ$ production, where $Z\to b\bar{b}$ is also reconstructed
in a single jet. The right hashed histogram shows the expected combinatorial background that is dominated by the $Z$+jets production. The solid black curve shows  the
final fit to the MC data. The dashed lines show each of the PDF components: the signal (blue),
$ZZ$ production (green), and the combinatorial background (red).}
\label{fig:mZHbb}
\end{center}
\end{figure}

In the search channel of the $ZH$ production mode, the boosted $H\to b\bar{b}$ is reconstructed in a single jet that is
based on the anti-$k_{\rm T}$ algorithm with a distance parameter $\Delta R=0.8$. We require  the jets to have 
$p_{\rm T}>300\,\gev$, $|\eta|<1.7$ and  $40\le\,m_{\rm jet}\le 200\,\gev$.
The $Z$ boson is reconstructed in the final states 
of $Z\to \ell\ell$, ($\ell = e,\mu)$ and $Z\to \nu\bar{\nu}$. Candidates of $Z\to \ell\ell$ decays are selected by combining 
an isolated, oppositely charged pair of electron or muon tracks and requiring the dilepton invariant
mass to be within $20\,\gev$ of the $Z$ boson mass. The leptons are also required to have  $p_{\rm T}>20\gev$ and $|\eta|<2.5$.
The identification of $Z\to \nu\bar{\nu}$ decays is done by requiring $E^{\rm miss}_{\rm T}\ge300\,\gev$ and 
$\Delta\phi(E^{\rm miss}_{\rm T},{\rm jet})>3$, where $\Delta\phi(E^{\rm miss}_{\rm T},{\rm jet})>3$ is the azimuthal angle between the
directions of the  $E^{\rm miss}_{\rm T}$ and the momentum of the selected $H$ jet candidate. The events with a $b$ jet that is not overlapping with the selected $H$ jet 
candidate are rejected. After applying  the above selection criteria, the dominant background left is the events from the SM $Z$+jets production,
where $Z\to\ell\ell, \nu\bar{\nu}$ and the recoiled jet is misidentified as a $H$ jet candidate. This background is greatly reduced by using the
BDT variable based on the double $b$-tagging and the jet substructure variables as described in Sec.~\ref{sec:tag_shape}. 
We optimize the  selection cut on the BDT variable by maximizing the signal significance of $S/\sqrt{B}$. 

The signal yield is extracted by a binned likelihood fit to the $m_{\rm jet}$ distribution of the selected $H$ jet candidates,
as shown in Fig.~\ref{fig:mZHbb}. The probability density functions (PDF) of the $H\to b\bar{b}$ and $Z\to b\bar{b}$ are
 modeled as two Gaussian functions.  The combinatorial background PDF is parametrized by a bifurcated Gaussian function that has 
 different widths on the left and right sides of the mean. The existence of the $Z\to b\bar{b}$ signal peak from the SM $ZZ$ production
 in this search channel provides an excellent calibration sample to constrain the $H\to b\bar{b}$ PDF parameters. In actual 
 data analysis at the LHC experiments, the parameters of the  $Z\to b\bar{b}$ PDF can be precisely determined from
 data by studying the boosted hadronically decaying $Z$ boson from the $Z$+jets production~\cite{Aad:2014haa}. 
 The background PDF can be also constrained using the events from the multijet production. In the default fit,
 the means of the Gaussian functions are allowed to float with a constant difference that is
fixed to the MC predicted mass difference between the $Z$ and Higgs bosons. The widths of two Gaussian functions are set to the values predicted by MC simulation.
 The mean of the bifurcated Gaussian function is allowed to be free in the fit, while the widths are fixed to the MC
predicted values. The fit result for the MC simulated events sample that is equivalent to $400\,\ifb$ of LHC data at the $14\,\tev$ center-of-mass 
energy is shown in Fig.~\ref{fig:mZHbb}. The fit yields more than $5\,\sigma$ of significance for both the $H\to b\bar{b}$ and $Z\to b\bar{b}$ signals.

\section{Conclusion}
In this paper we study the identification of a bottom quark-antiquark pair inside
in a single jet with high transverse momentum  by using the jet substructure in the center-of-rest frame of the jet. 
We demonstrate that the method can significantly reduce the QCD jet background while maintaining a high 
identification efficiency of the boosted Higgs boson decaying to a $b\bar{b}$ pair even under
a very large pileup condition. The study  shows a good prospective on searches for 
$H\to b\bar{b}$ decay in the $VH$ production mode for the LHC experiments at the $14\,\tev$ center-of-mass
energy, and it is  complementary to the existing searches~\cite{Chatrchyan:2013zna,Aad:2014xzb}
in which each of the $b$ quarks decayed from the Higgs boson is reconstructed as an individual jet.
The proposed technique can be also used to search for new physics phenomena beyond the SM, such as
possible dark matter candidates 
produced in association with the SM Higgs boson~\cite{Carpenter:2013xra}.

\label{sec:conclusion}

\vspace{0.5cm}
\section{Acknowledgments}
We thank Jim Cochran and Soeren Prell for many discussions. 
This work is supported by the Office of Science of the U.S. Department of
Energy under Contract No. DE-FG02-13ER42027.

\appendix 

\section{Jet substructure distribution}
\label{appx:1}
\begin{figure}[!htb]
\begin{center}
\includegraphics[width=0.23\textwidth]{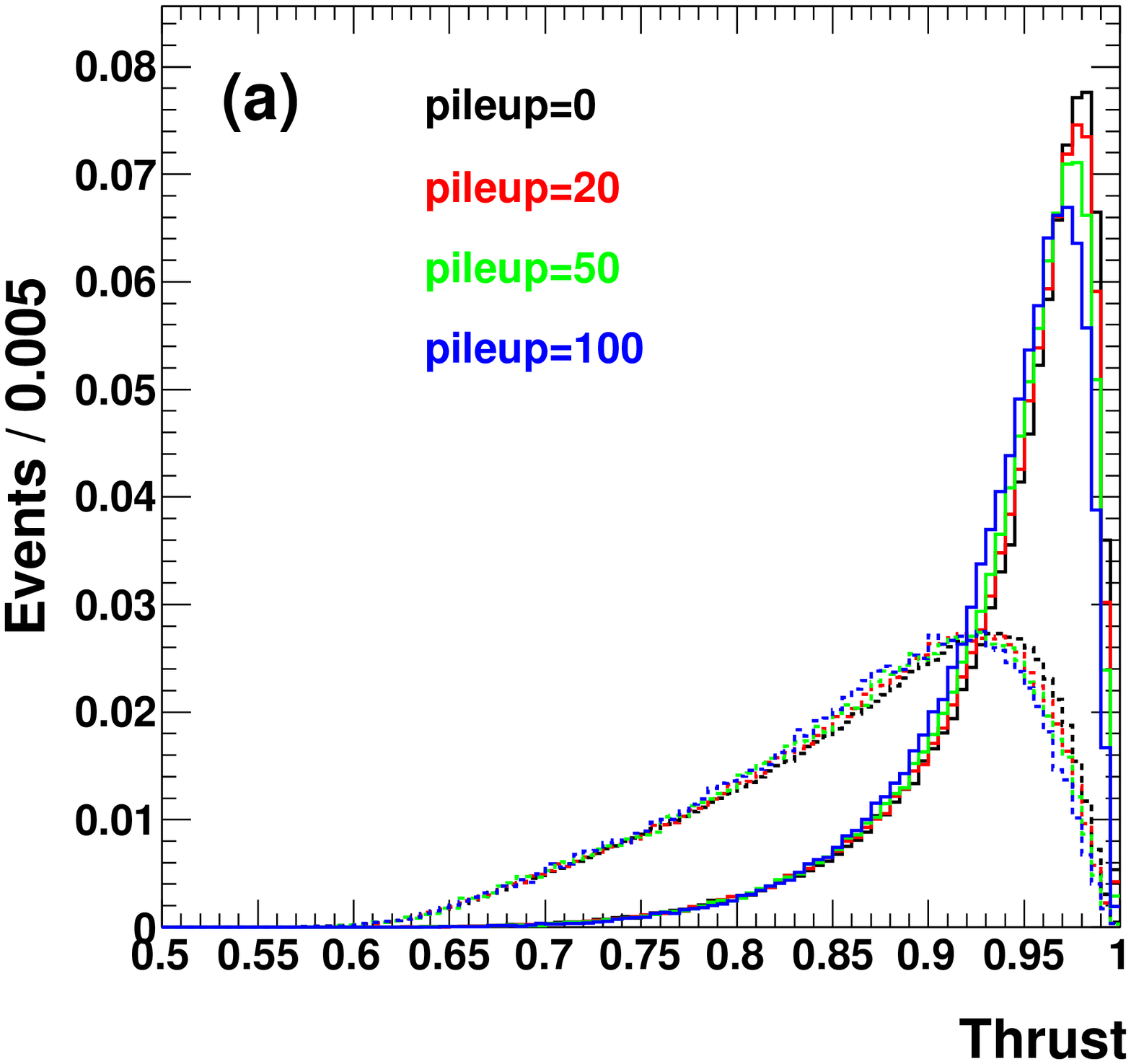}
\includegraphics[width=0.23\textwidth]{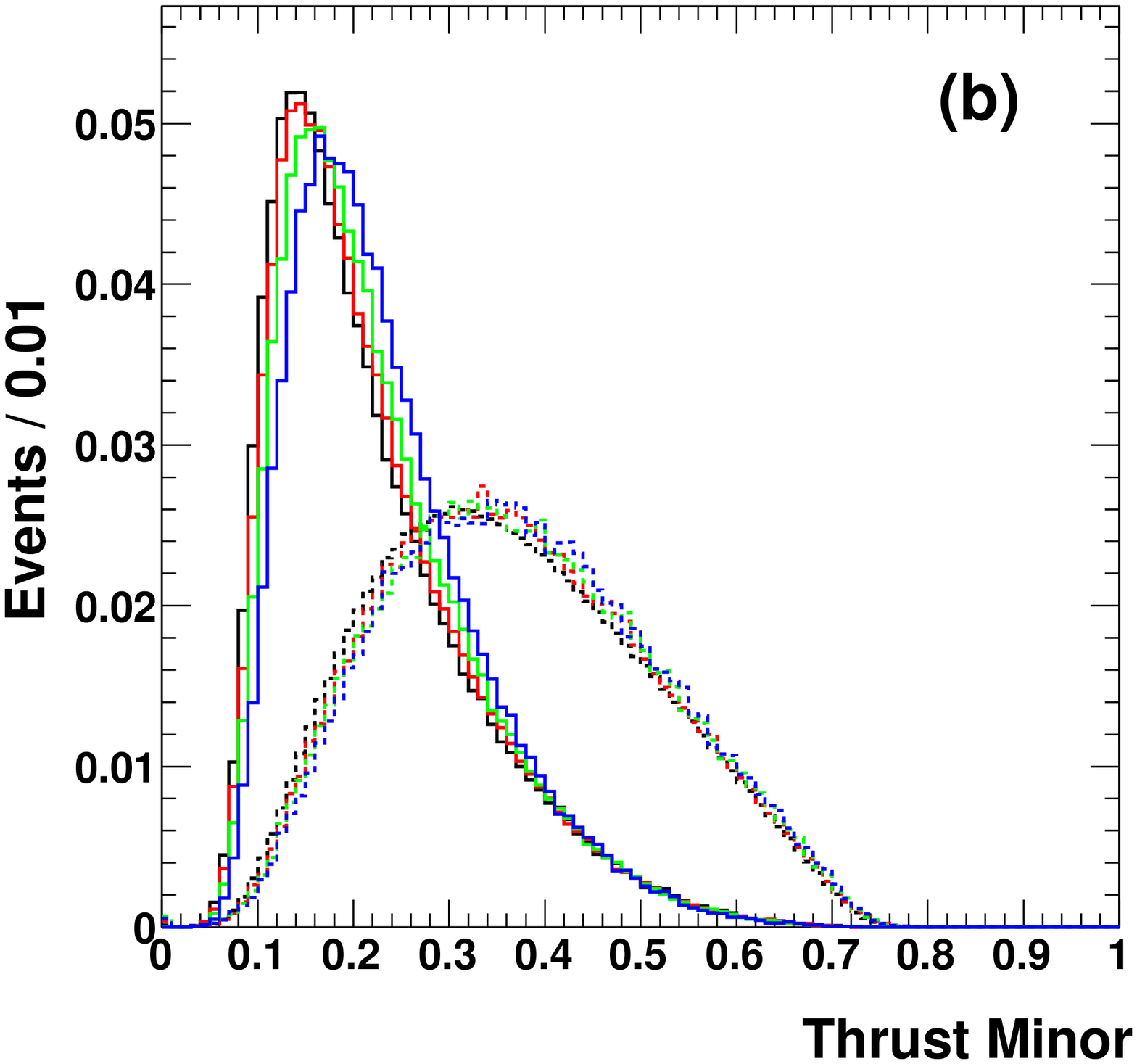}
\includegraphics[width=0.23\textwidth]{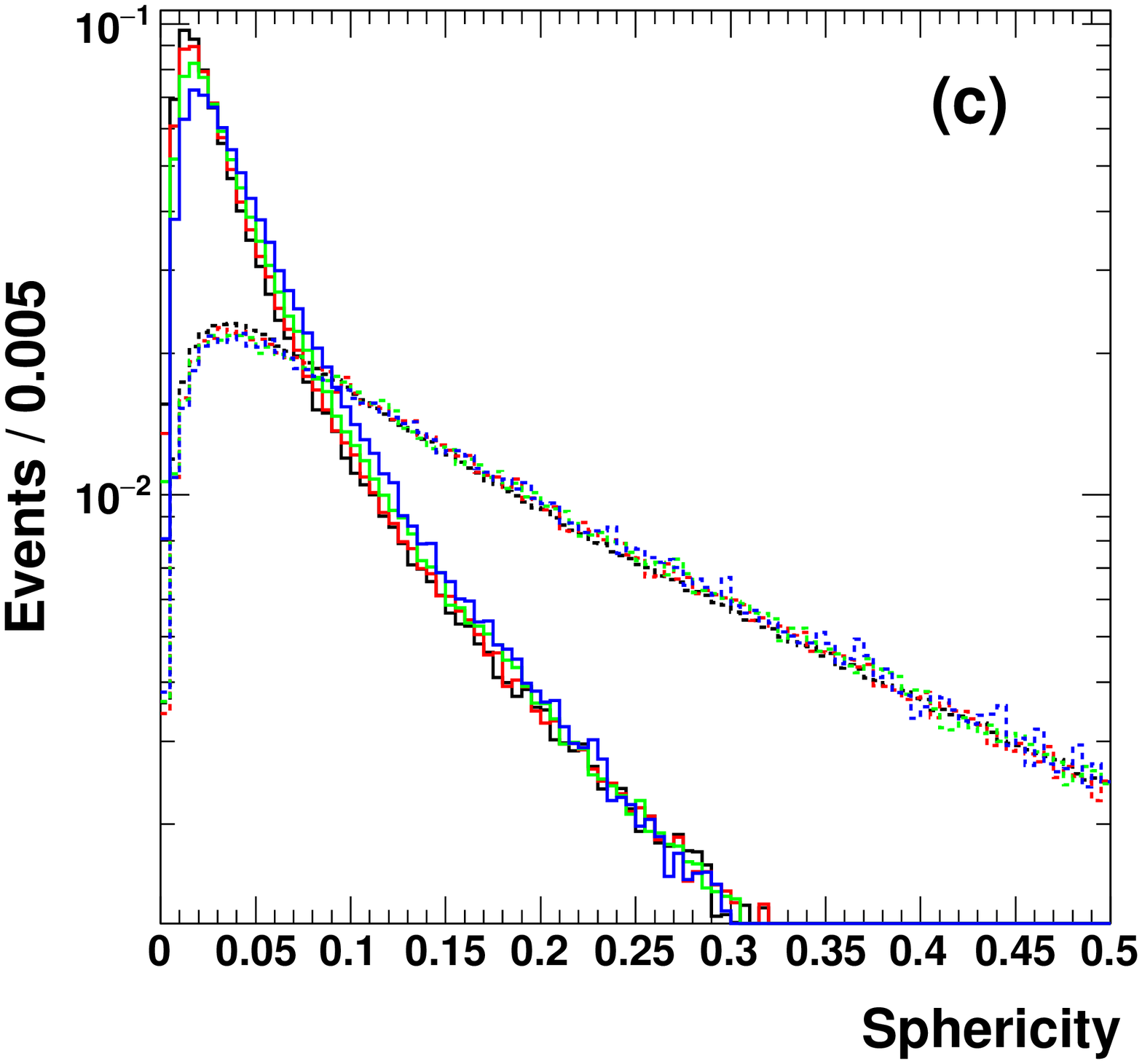}
\includegraphics[width=0.23\textwidth]{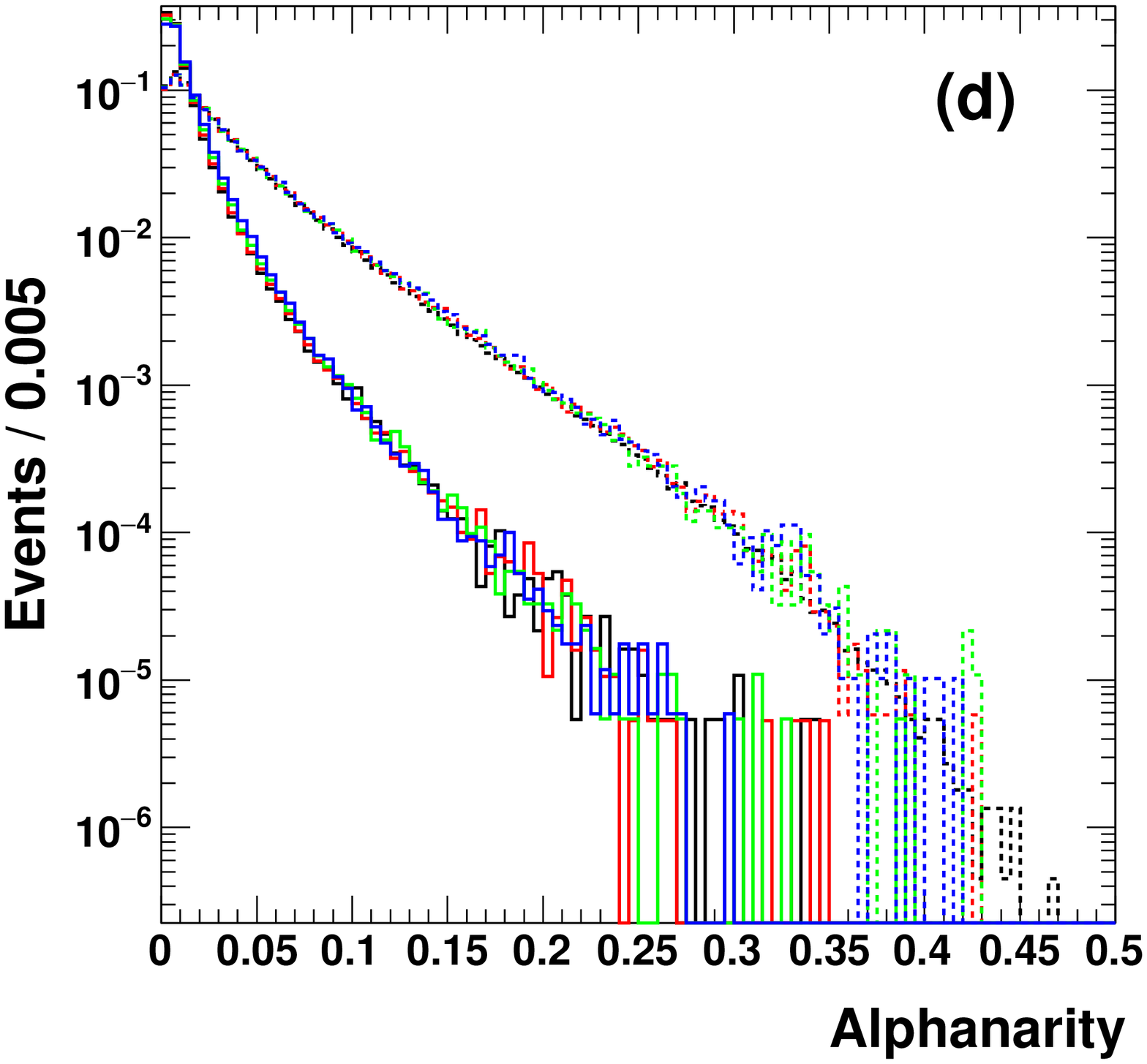}
\includegraphics[width=0.23\textwidth]{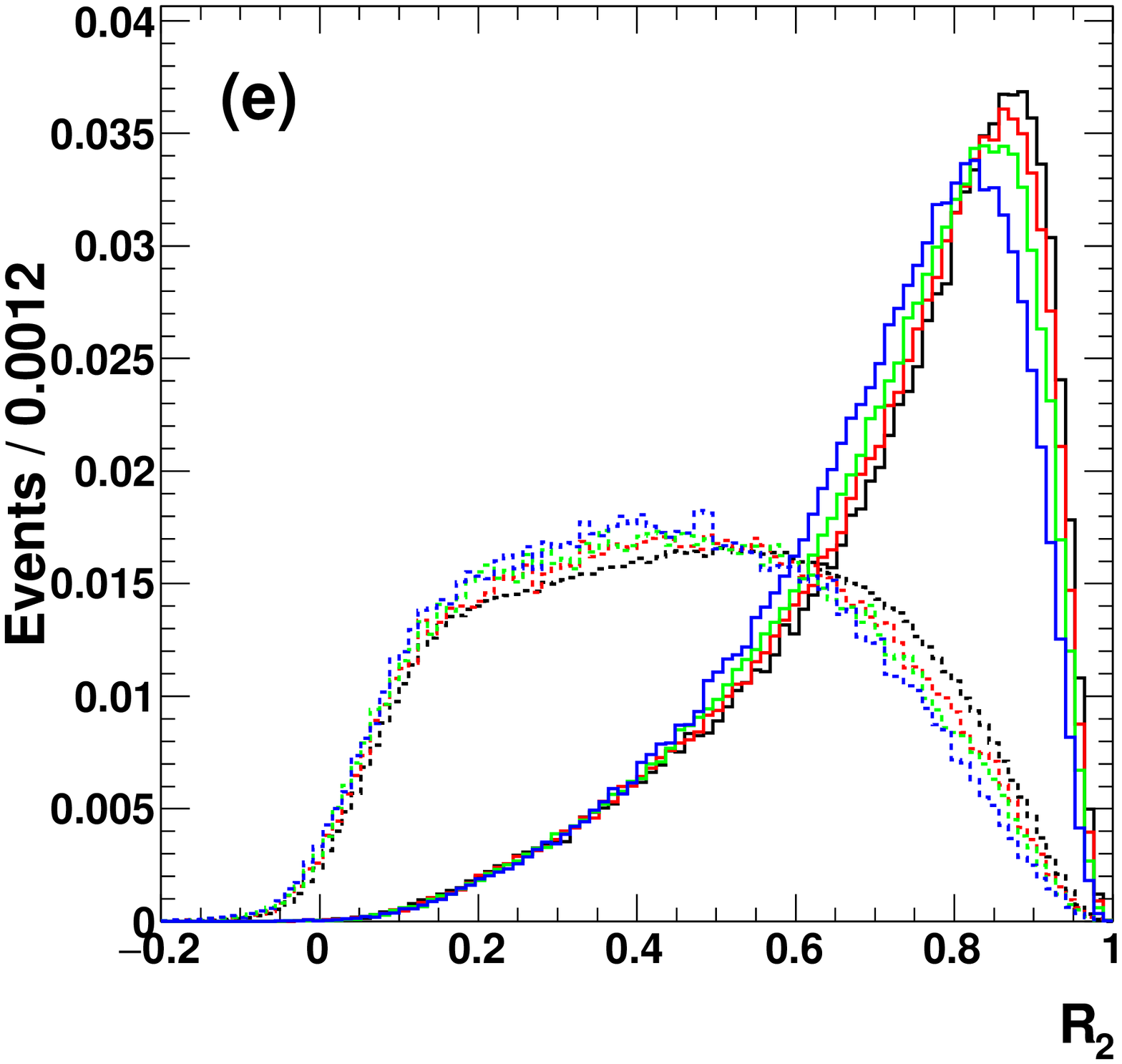}
\includegraphics[width=0.23\textwidth]{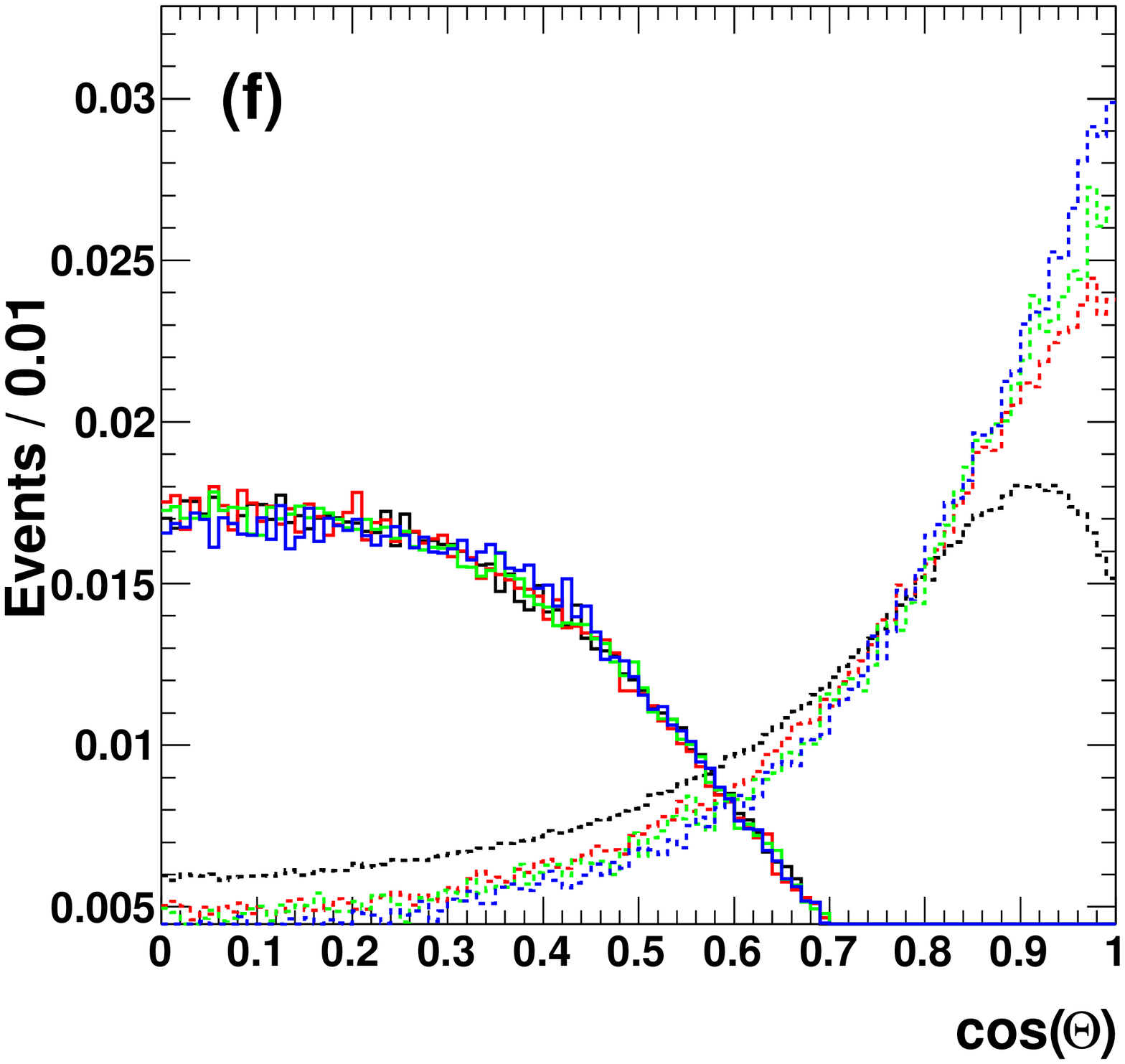}
\caption{The distribution of the jet shape variables in the center-of-mass frame of the jet:
(a) thrust, (b) thrust-minor, (c) sphericity, (d) aplanarity,
(e) $R_2$, and (f) $\cos\Theta$ for the $H$ jet signal and QCD jet background.
All the distributions are normalized to unity.}
\label{app:jetsubstructure}
\end{center}
\end{figure}

\end{document}